# Issues In Disintermediation In The Real Estate Brokerage Sector.

**Michael C. Nwogugu**
Address: Enugu 400007, Enugu State, Nigeria.
Email: Mcn2225@aol.com; mcn2225@gmail.com.
Phone: 234 814 906 2100.

**Abstract**[1].
This article introduces new models of dis-intermediation of the real estate broker by the buyer or the seller. The decision to retain a real estate broker is critical in the property purchase/sale process. The existing literature does not contain analysis of: 1) information asymmetry, 2) the conditions under which it will be optimal to disintermediate the broker, 3) social capital and reputation, 4) the impact of different types of real estate brokerage contracts. The article shows that dis-intermediation of the real estate broker by the seller or buyer may be optimal in certain conditions.

## 1. Internet-Based Dis-Intermediation In Real Estate Transactions.

In the US, the residential real estate brokerage industry alone has annual revenues of more than $60 billion. However, the industry is consolidating and brokerage commissions are declining. The industry is regulated. Many companies have tried (mostly un-successfully to shift brokerage operations onto the Internet. These companies have learned to use the Internet as a marketing, information-resource and educational tool. Benjamin, Chinloy, Judd & Winkler (2005); Tse & Webb (2002); Bond & Seiler, Blake (2000); Diehl, Kornish & Lynch (2003); Bolton, Warlop & Alba (2003); Carmon & Ariely (2001); Stavrovski (2004); Wu & Colwell (1986); Katz (1990); Bakos, Lucas, Oh & Simon, et al (2005).

## 2. Existing Literature.

The existing literature on the economics of real estate brokerage, provision of online information, and possible disintermediation is extensive. See: Gwin (2004); Miceli, Pancak & Sirmans (2000); Yavas, Miceli & Sirmans (2001); Tuccillo (1997); Curran & Schrag (2000); Rutherford, Springer &Yavas (2004); Muhanna & Wolf (2002);

---

[1] A later version of this article was published as:
Nwogugu, M. (2007). Issues In Disintermediation In The Real Estate Brokerage Industry. *Applied Mathematics &*

D'Urso (_____); Johnson, Springer & Brockman (2005); Quint & Wako (2004); Lurie (2004); Schmitz (2000); Scott (2000); Munneke & Yavas (2001); Mantrala & Zabel (1995); Anglin (1997); Sirmans, Turnbull & Benjamin (1991); Zumpano & Elder (1994); Zumpano, Elder & Baryla (1996); Wheaton (1990); Bajtelsmit & Worzala (1997); Yavas (1992); Arnold (1999). On dynamical systems, also see: Beer (2000); Dellnitz & Junge (1999); Friedman & Sandler (1996); Agarwal, Bohner, O'Regan & Peterson (2002); Iacus (2001); Van Gelder (1998); Izmailov & Solodov (2001); Iri (1997); Mordukhovich & Shao (1997); Van Dalen (June 1995); Anglin & Arnott (1991); Hodgkinson (1997); Williams (1998); Korczynski & Ott (2005); Garmaise & Moskowitz (2004); Kauffman, Subramani & Wood (2000); Elder (2000); Bui, Yen & Sankaran (2001); Rutherford, Springer & Yavas (2004); Dale-Johnson (1998); Dolde & Tirtiroglu (1997); Fleck (2000); Evans (1998); Janssen & Jobson (1980); Moore (1991); Nelles (2002); Quan (2002); Williams (1998); Yavas & Colwell (1999); Benjamin, Chinloy, Jud & Winkler (2006); Lewis & Anderson (1999); Geltner, Kluger & Miller (1991); Delcoure & Miller (2002); Angilini (2005); Williams (1998); Anderson & Fok (1998); Anderson, Lewis & Zumpano (1999); Austin (1973); Barlett (1981); Benjamin, Jud & Sirmans (2000); Guttery, Baen & Benjamin (2000); Jud, Rogers & Crellin (1994); Jud & Winkler (1998); Jud & Winkler (2000); Lewis & Anderson (1999); Marsh & Zumpano (1988); Miller & Shedd (1979); Muhanna (2000); Muneke & Yavas (2001); Owen & Kickbacks (1977); Turnbull (1996); Worzala & McCarthy (2006); Wachter (1987); Yavas (2001); Yinger (1993); Zumpano, Anderson, Baryla & Johnson (2001); Zumpano, Elder, Elder & Crellin (1993). On reputation and social capital, see: Chemmanur & Fulghieri (1994); Fang (2005); Bouzdine & Bourakova-Lorgnier (April 2004); Guennif & Revest (2005); Black, Carnes & Richardson (2000); Swain (2003); Al-Ubaydli (March 2005); Burt (2000); Seierup (1996); Fine (1999); Annen (_____); Schmid (2000); Glaeser, Laibson & Sacedote (2002); Narayan & Pritchett (1999); Durlauf (2002); Bowles & Gintis (2002); Dasgupta (2005); Milgrom & Roberts (1982); Kreps & Wilson (1982).

However, the existing literature omits the following analysis:

- **T**he amount of information that the real estate Broker should provide on the web.
- **W**hen the Broker should provide information on the web.
- Relevant conditions under which the buyer and or seller should dis-intermediate the real estate broker; and quantification of such conditions.

---

*Computation,* 186(2), 1054-1064.

- Quantification of effects of the type of real estate brokerage contract; and constraints imposed by the brokerage contracts.
- Differentiation among different types of customer search costs (physical and Internet).
- The buyer's/seller's decision to retain or dis-intermediate a real estate Broker within the context of the buyer's and seller's Social Capital and or Reputation Capital.
- Effect of Broker's Social Capital and Reputation Capital.
- Incorporation/recognition of limitations of Rationality in economic analysis/modeling - brokers, buyers and sellers are subject to emotions, regret, risk-aversion, and other psychological effects.
- Analysis of buyer utility gained from interactions with Broker; analysis of seller utility gained from dealing with the Broker. Most studies errornuously equate buyer/seller utility with monetary outcomes for the buyer/seller.
- Elimination of 'equilibrium' in economic analysis; Colman (2003); Hertwig & Ortmann (2001); Arthur (April 1999); Bromiley & Papenhausen (______).

### 3. Models.

This section develops theoretical models of real estate brokerage, information asymmetry among the broker, buyer and seller; broker supply of information via the Internet, and possible disintermediation of the broker. The models result in testable hypothesis about:

- **W**hen the Broker should provide information on the web. **T**he optimal amount of information that the Broker should provide on the web.
- **W**hen the buyer should dis-intermediate the broker.
- Conditions under which the seller should disintermediate the broker
- The Broker's objective function.
- The effects of social capital and reputation capital on the decision to disintermediate or hire the broker.

*Assumptions*:
- Brokers do not restrict access to information that they provide on the Internet.

- The only difference in Closing Costs between situations where the broker is used and where the broker is dis-intermediated, is the Broker's fees.

**P** = estimated/appraised value of housing unit. $P \in (0, +\infty)$.
**$P_b$** = The value of the housing unit to the Buyer. $P_b \in (0, +\infty)$.
**$P_s$** = Value of home to the Seller. $P_s \in (-\infty, +\infty)$.
**c** = percentage commission rate paid to Broker by Seller. $0 < c < 1$.
**$B_b$** = broker's total fixed cost of providing information to a buyer – this includes rent, utilities, subscriptions to databases, telephones, administrative staff, operating expenses, etc.. Most are fixed costs and semi-fixed costs. $B_b \in (-\infty, +\infty)$.
**$B_n$** = Broker's total cost of searching for a new client. Assumes that broker has to prospect *n* number of prospective clients for each actual new client obtained. $B_n \in (-\infty, +\infty)$.
**$B_{op}$** = Broker's total operating expenses per transaction. Assumes that broker has to prospect *n* number of prospective clients for each actual new client obtained. $B_{op} \in (-\infty, +\infty)$.
**$B_s$** = Broker's fixed cost of dealing with seller and listing property. $B_s \in (-\infty, +\infty)$.
**$B_i$** = Broker's amortized periodic fixed cost of providing information on the website – this includes the amortized cost of building the website, hosting/bandwidth costs, website maintainance costs and the incremental costs of uploading info0rmation about any one property. However, the incremental costs of uploading information about any one property are relatively small. $B_i \in (-\infty, +\infty)$.
**$B_{it}$** = Broker's cost of providing information on the website – this includes the present value of the total cost of building the website, hosting/bandwidth costs, website maintainance costs and the incremental costs of uploading information about any one property. $B_{it} \in (-\infty, +\infty)$.

**I** = the amount of information (I) communicated to the buyer by broker; and the following conditions apply:
1. $\partial I/\partial B_b > 0$; $\partial^2 I/\partial B_b^2 > 0$.
2. $I = I_p + I_i$

**$I_p$** = value of information provided personally by broker. $I_p \in (-\infty, +\infty)$.
**$I_i$** = value of information provided by broker on the website. $I_i \in (-\infty, +\infty)$.
**$I_o$** = value of information (about subject property and market conditions) obtained from all websites (other websites, and Broker's website) by buyer. $I_i \in I_o$. $I_o \in (-\infty, +\infty)$.
**$\psi_b$** = Buyer's search costs without internet. $\psi_b \in (-\infty, +\infty)$.
**$\psi_{bi}$** = Buyer's search costs with internet use. $\psi_{bi} \in (-\infty, +\infty)$.
**$\psi_s$** = Seller's search costs with Broker but without internet and other resources. $\psi_s \in (-\infty, +\infty)$.
**$\psi_{si}$** = Seller's search costs with internet use and other resources but without Broker. $\psi_{si} \in (-\infty, +\infty)$.
**$\psi_{sb}$** = Seller's search costs with Broker and with internet use. $\psi_{sb} \in (-\infty, +\infty)$.
**$U_{ip}$** = utility gained by buyer from information that broker communicates personally – utility measured in terms of usefulness towards closing the transaction, psychological comfort, attention to detail, enhancement of ability to compare/contrast properties, perception of neighborhood, etc.. Utility is different from value of information (which strictly refers to usefulness towards closing a transaction). $U_{ip} \in (-\infty, +\infty)$.
**$U_{iw}$** = utility gained by buyer from information that broker communicates via the web – utility measured in terms of usefulness of information towards closing the transaction, choice of properties to buy, psychological comfort, attention to detail, enhancement of ability to compare/contrast properties, perception of neighborhood, etc.. Utility is different from value of information (which strictly refers to usefulness towards closing a transaction). $U_{iw} \in (-\infty, +\infty)$.
**$U_a$** = utility gained by buyer from information he/she obtains from the web (including the broker's website) without communicating with the Broker. $U_a \in (-\infty, +\infty)$.
**$U_{sp}$** = utility gained by seller from information that broker communicates personally – utility measured in terms of usefulness towards closing the transaction, psychological comfort, attention to detail, enhancement of ability to compare/contrast properties, perception of neighborhood, etc.. Utility is different from value of information (which strictly refers to usefulness towards closing a transaction). $U_{sp} \in (-\infty, +\infty)$.

$U_{sw}$ = utility gained by seller from information that broker communicates via the web – utility measured in terms of usefulness of information towards closing the transaction, choice of properties to buy, psychological comfort, attention to detail, enhancement of ability to compare/contrast properties, perception of neighborhood, etc.. Utility is different from value of information (which strictly refers to usefulness towards closing a transaction). $U_{sw} \in (-\infty, +\infty)$.

$U_{sa}$ = utility gained by seller from information he/she obtains from the web (including the broker's website) without communicating with the Broker. $U_{sa} \in (-\infty, +\infty)$.

$\pi_b$ = additional closing costs, other than broker's labor-rate, if broker is used. This includes Seller's time spent negotiating/discussing terms with Broker. $\pi_b \in (-\infty, +\infty)$.

$\pi_i$ = additional closing costs, other than broker's labor-rate, if broker is dis-intermediated. $\pi_i \in (-\infty, +\infty)$.

$\pi_{sb}$ = closing costs, if broker is used. $\pi_{sb} \in (-\infty, +\infty)$.

$\pi_s$ = closing costs, if broker is not used. $\pi_s \in (-\infty, +\infty)$.

$E_s$ = a state where broker has a super-exclusive listing – seller must pay broker regardless of how buyer meets seller. Broker's profits are guaranteed upon sale. Broker can share commissions with other brokers who participate in the sale. The state is characterized and quantified based on the Broker's effort (hours spent, value of broker's mental processes), Broker's Labor-rate, and value of brokerage contract. $E_s \in (-\infty, +\infty)$.

$E_p$ = a state where broker has semi-exclusive listing and broker is paid only if the transaction is consumated as a direct result of a buyer introduced by the broker. Seller can still sell based on his/her own efforts and not pay the commission. The state is characterized and quantified based on the Broker's effort (hours spent, value of broker's mental processes), Broker's Labor-rate, and value of brokerage contract. $E_p \in (-\infty, +\infty)$.

$E_m$ = a state where multiple listing is used, and broker is only one of many brokers that can sell the property. Broker. Broker is paid only if the transaction is consumated as a direct result of a buyer introduced by the broker. The state is characterized and quantified based on the Broker's effort (hours spent, value of broker's mental processes), Broker's Labor-rate, and value of brokerage contract. $E_m \in (-\infty, +\infty)$.

$\rho_p$ = probability of closing the sale if broker provides information primarily in physical space, and the transaction is done only in (or primarily in) physical space. Probability is assessed in terms of broker's track record, average time-on-the-market, local demographics, broker's social capital and internet networks, broker's reputation, average weekly leisure time (#of hours) of males/females in the area with sufficient income to purchase, etc.

$\rho_i$ = probability of closing the sale if information is provided by broker primarily through the Internet, and transaction is processed mostly through the Internet (online forms, automated mortgage applications, etc.). Probability is assessed in terms of broker's track record, average time-on-the-market for properties in the area, broker's social capital and internet networks, broker's reputation, local demographics, internet penetration rate, internet access, education levels, etc.

$\rho_s$ = probability of closing the sale if seller does not hire the broker. The Seller uses the Internet and other resources to sell the property. Probability is assessed in terms of average time-on-the-market for housing units in the area, local demographics, internet penetration rate, internet access, education levels, seller's knowledge of the internet, seller's knowledge of local real estate markets and mortgages, seller's disposable income, seller's social capital, seller's reputation capital, existence of un-usual circumstances (healthcare problems, seller relocation, seller financial distress, etc.).

$ά$ = broker's effort – which is the sum of the broker's labor/time costs, office operating expenses (rent, equipment, professional fees, transportation costs, office supplies, cost of memberships in professional organizations and MLC systems, professional insurance, etc.), and other costs required to complete one purchase/sale transaction. $ά \in (-\infty, +\infty)$.

$ά_s$ = buyer's perception of value of broker's effort. $ά_s \in (-\infty, +\infty)$.

$RC_{br}$ = Broker's Reputation capital gained from the transaction– which is non-cash and varies with broker's efforts, broker's knowledge, transaction sizes, etc. and $-\infty < \partial RC_{br}/\partial e < \infty$; $RC_{br} \in (-\infty, +\infty)$.

$SC_{br}$ = Broker's social capital gained/lost due to the transaction – which is non-cash and varies with broker effort, knowledge, ability to build and maintain relationships, social networks, etc., and $-\infty < \partial SC_{br}/\partial e < \infty$; $SC_{br} \in (-\infty, +\infty)$.

$SC_s$ = value of Seller's social capital. Seller's social capital is a function of familiarity with the neighborhood/town, social networks, reputation, etc. $SC_s \in (-\infty, +\infty)$.

$SC_b$ = value of Buyer's social capital. $-\infty < SC_b < \infty$. Seller's social capital is a function of familiarity with the neighborhood/town, social networks, reputation, etc. $SC_b \in (-\infty, +\infty)$.

A buyer will disintermediate the broker if the following conditions exist:
1. $I_i > I_p \mid (U_{iw} > U_{ip})$; $(I_i + I_o) > I_p \mid (U_{iw} > U_{ip})$;
2. $I_i \approx \psi_b$; information costs are similar
3. $[\{cP + \psi_b + \pi_b\} \mid \{\{P_s \approx P_b\} \cap \text{Max}(E_m, E_p, E_s)\}] > \{\psi_{bi} + \pi i + U_{iw}\}$
4. $[\{cP + \psi_b + \pi_b\} \mid \{(U_{iw} > U_{ip}) \cap \text{Max}(E_m, E_p, E_s)\}] > \{\psi_{bi} + \pi i + U_{iw}\}$
5. $\psi_b > \psi_{bi}$; and $U_{iw} > U_{ip}$
6. $(\partial^2\psi_b/\partial U_{ip}^2) < \text{Min}[(\partial^2\psi_{bi}/\partial U_{iw}^2), 1]$; and $(\partial\psi_b/\partial U_{ip}) < \text{Min}[(\partial\psi_{bi}/\partial U_{iw}), 1]$;
7. $(\partial U_{iw}/\partial\pi_i) > \text{Min}(\partial U_{ip}/\partial\pi_b), 0]$; and $\{(\partial^2 U_{iw}/\partial\pi_i^2) > \text{Min}[(\partial^2 U_{ip}/\partial\pi_b^2), 0]\}$;
8. $(\partial^2 I_o/\partial\psi_{bi}^2) > \text{Max}[\partial^2(I_p+I_i)/\partial\psi_b^2, 1]$; and $(\partial I_o/\partial\psi_{bi}) > \text{Max}(\partial(I_p + I_i)/\partial\psi_b), 1]$
9. $\partial(I_p+I_i)/\partial\text{Max}(E_m, E_p, E_{sw}) < 1$;
10. $\partial(I_p+I_i)/\partial(U_{ip}+U_{iw}) < \text{Min}[(\partial I_o/\partial U_a), 1]$; and $\partial^2(I_p+I_i)/\partial(U_{ip}+U_{iw})^2 < \text{Min}[(\partial^2 I_o/\partial U_a^2), 1]$;
11. $\partial^3(I_p+I_i)/\partial(U_{ip}+U_{iw})^3 < 1$; and $\partial^3 I_o/\partial U_a^3 < 1$;
12. $(\partial^3 P_b/\partial P^3) > \text{Max}[(\partial^3 I_o/\partial(I_p+I_i)^3), 1]$; and $\{\partial P_b/\partial P > \text{Max}[(\partial I_o/\partial(I_p+I_i)), 1]\}$;
13. $(\acute{\alpha}_s) < (cP + \pi_{sb} + I_p + I_i)$
14. $[\partial(\acute{\alpha}_s)/\partial(\pi_{sb} + I_p + I_i)] < 1$
15. $(SC_b - \psi_{bi} - \psi_b) > (U_{ip} + U_{iw} + I_p + I_i + \pi_b)$
16. $\partial SC_b/\partial\text{Max}(\psi_{bi},\psi_b) > \text{Max}[1, \partial(U_{ip}+U_{iw})/\partial(I_p+I_i+\pi_b)$; and
17. $\partial^2 SC_b/\partial\text{Max}(\psi_{bi}, \psi_b)^2 > \text{Max}[0, \partial^2(U_{ip}+U_{iw})/\partial(I_p+I_i+\pi_b)^2$;
18. $\partial SC_b/\partial\text{Max}(\psi_{bi}, \psi_b) > \text{Max}[1, \partial(\rho_i \cap \rho_p)/\partial(U_{ip}+U_{iw})]$;
19. $\partial^2 SC_b/\partial\text{Max}(\psi_{bi}, \psi_b)^2 > \text{Max}[0, \partial^2(\rho_i \cap \rho_p)/\partial(U_{ip}+U_{iw})^2]$;

The broker will have incentives to provide information on the website if the following conditions exist:
1. $[\{(cP*\rho_p) - B_b - B_s\} < \{B_i\}] \mid [(\psi_b > \psi_{bi}) \cap E_p]$
2. $[U_{ip} < U_{iw}] \mid \text{Max}(E_s, E_p)$
3. $(\psi_{bi} * \rho_i) > \{cP * \rho_p\}$
4. $\partial\rho_i/\partial\rho_p \approx 0$
5. $\partial\rho_i/\partial B \approx 0$; $\partial\rho_p/\partial B \approx 0$
6. $\partial B_i < \partial(RC_{br} + SC_{br})$;
7. $\partial(RC_{br}+SC_{br})/\partial B_i > 1$

The Seller will dis-intermediate the Broker if the following conditions exist:
1. $\rho_s > (\rho_p \cap \rho_i)$; . $\rho_s > \rho_p, \rho_i$;
2. $\partial I_o/\partial\psi_{si} > \text{Max}[\partial(I_p+I_o)/\partial\psi_{sb}, 1]$; and $\partial^3 I_o/\partial\psi_{si}^3 > \text{Max}[\partial^3(I_p+I_o)/\partial\psi_{sb}^3, 1]$;
3. $\partial U_a/\partial\psi_{si} > [\partial(U_{sp}+U_{sw})/\partial\psi_s, 1]$; and $\partial^3 U_a/\partial\psi_{si}^3 > [\partial^3(U_{sp}+U_{sw})/\partial\psi_s^3, 1]$
4. $\psi_s > \psi_{si}$; and $\psi_s \mid \text{Max}(E_m, E_p, E_s) > \psi_{si} \mid \text{Max}(E_m, E_p, E_s)$
5. $I_o > (I_i + I_p)$; and $I_o \mid \text{Max}(E_m, E_p, E_s) > (I_i+I_p) \mid \text{Max}(E_m, E_p, E_s)$
6. $\partial P_s/\partial P > 1$;
7. $\partial\pi_{sb}/\partial(U_{sp}+U_{sw}) > \partial\pi_s/\partial U_{sa}$; and $\pi_{sb} > \pi_s$;
8. $\partial P/\partial\pi_{sb} > \text{Max}[\partial P > \partial\pi_s, 1]$; $\partial P_s/\partial\pi_{sb} > \text{Max}[\partial P_s > \partial\pi_s, 1]$; $\partial P_s/\partial P > 1$; $\partial P/\partial c > 1$
9. $(\psi_{sb}+\pi_{sb}) > (\psi_{si}+\pi_s)$; and $\partial(\psi_{sb}+\pi_{sb})/\partial P_s > \partial(\psi_{si}+\pi_s)/\partial P_s$;
10. $\partial(\psi_{sb}+\pi_{sb})/\partial c > \partial(\psi_{si}+\pi_s)/\partial c$
11. $\partial(\psi_{sb}+\pi_{sb})/\partial\pi_{sb} > \partial(\psi_{si}+\pi_s)/\partial\pi_{sb}$
12. $\rho_s > \rho_p, \rho_i$
13. $\int_0^T \{\rho_s (P_s - \pi_b - \pi_s - \psi_{si}) \partial t > \text{Max}[\int_0^T \{(\rho_p)*(P - \pi_b - \pi_{sb} - \psi_s)\partial t\}, \int_0^T \{(\rho_i \cap \rho_p)*(P - \pi_b - \pi_{sb} - \psi_{sb}) \partial t\}]$
14. $(SC_s - \psi_{si} - (\pi_{sb}.\pi_{sb})) > (U_{sw} + U_{sp} + I_p + I_i + \pi_{sb})$
15. $\partial SC_s/\partial\text{Max}(\psi_{si},\psi_s) > \text{Max}[1, \partial(U_{sp}+U_{sw})/\partial(I_p+I_i+\pi_b)$;
16. $\partial^2 SC_s/\partial\text{Max}(\psi_{si}, \psi_s)^2 > \text{Max}[0, \partial^2(U_{sp}+U_{sw})/\partial(I_p+I_i+\pi_b)^2$;

17. $\partial SC_s/\partial \text{Max}(\psi_{si}, \psi_s) > \text{Max}[1, \partial(\rho_i \cap \rho_p)/\partial(U_{sp}+U_{sw})]$;
18. $\partial^2 SC_s/\partial \text{Max}(\psi_{si}, \psi_s)^2 > \text{Max}[0, \partial^2(\rho_i \cap \rho_p)/\partial(U_{sp}+U_{sw})^2]$;

Regardless of whatever effort ($\acute{\alpha}$) broker makes, the broker's compensation consists of: **1)** the brokerage fee, which is cash compensation and does not vary with broker effort ($\acute{\alpha}$), and **2)** Reputation capital ($RC_{br}$) which is non-cash and varies with broker's efforts and knowledge, size of the transaction, frequency of broker's transactions, and the ratio of the broker's closed sale/purchase transactions to total the broker's possible sales/purchase transactions - and $\partial RC_{br}/\partial \acute{\alpha} \in (-\infty, +\infty)$; and **3)** social capital ($SC_{br}$) which is non-cash and varies with broker effort, broker's knowledge, broker's ability to build and maintain relationships, social networks, etc., and $\partial SC_b/\partial \acute{\alpha} \in (-\infty, +\infty)$. The literature on social capital, trust and reputation capital is extensive. See: Rangan (2000); Lee, Wu, Maguluru & Nichols (2006); Kauffman, Subramani & Wood (2000); Garmaise & Moskowitz (2004); Ferrary (2003); Korczynski & Ott (2005); Pollock (2004). Kreps & Wilson (1982); Milgrom & Roberts (1982); Hellman & Murdock (1998), On reputation and social capital, see: Chemmanur & Fulghieri (1994); Fang (2005); Bouzdine & Bourakova-Lorgnier (April 2004); Guennif & Revest (2005); Black, Carnes & Richardson (2000); Swain (2003); Al-Ubaydli (March 2005); Burt (2000); Seierup (1996); Fine (1999); Annen (_____); Schmid (2000); Glaeser, Laibson & Sacedote (2002); Narayan & Pritchett (1999); Durlauf (2002); Bowles & Gintis (2002); Dasgupta (2005); Milgrom & Roberts (1982); Kreps & Wilson (1982). The typically constant brokerage commission *cP*, is modified by the possible states $E_s$, $E_m$ and $E_p$, and quality of information provided by the broker, $I_i$ and $I_p$. The broker's effort levels are constrained by the Brokerage Agreement, the agreed upon brokerage fee, state contract laws, and state real estate brokerage laws, the Broker's capital, the Broker's opportunity costs, and the mechanics of Multiple Listing Service systems. Hence, one of the broker's objectives is to ensure that the sum of the value of his/her effort plus other resources contributed to the search, does not exceed the brokerage fee less a reasonable profit.

Contrary to most housing demand models and brokerage industry models, the occurrence of the event '*cP'* isn't binary (ie. the housing-unit/property is sold or is not sold), and there are some key considerations:

- There are psychological influences that affect the buyer's decision and the sellers decision – including perceived sincereity of broker, affinity for certain locations/features/sizes/colors, time considerations, perception/cognition, intuition about prices, the perceived financial burden arising from payment terms (size of down payment, etc.), and other factors.

- The buyer's choice of, or combinations of $I_i$ and $I_p$ is highly subjective and depends on many variables (including region, internet access, level of education of buyer, buyer's perception/cognition, buyer's income, family structure, buyer's temperament, value of buyer's time, buyer's familiarity with internet transactions, etc.) and is not completely controlled by the broker – the buyer can search for information about properties offered by the broker by using other databases and website, and even if the buyer does not have the subject-property's address, the buyer can also easily do online searches for comparable properties in the same region, and gain high-value information about prices and sale terms.

- In calculating how much information that the broker should provide on the web, the following considerations apply:

**a)** In state $E_m$, the buyer can use the information and still close the transaction with another broker.

**b)** In most instances, $I_p > I_i$; and the incremental cost of providing Ii is relatively small. A broker's website can be built for $400-3,000 and website hosting costs are $1-30 per month. The seller will typically supply pictures of the property and input property-specific data. The broker can easily obtain neighborhood data from commercial databases.

**c)** If the broker chooses to provide information online, the broker will typically provide standard data about the property or a minimum amount of information. The key issue then is which information most influences property sales/purchases and is discretionary ? the answer is most likely, comprehensive pictures, historical details of significance (previous owners/tenants, previous uses, etc.), construction details, and information about property conditions. For the broker, the incremental costs of providing such information is relatively minimal because most of it is provided by the seller. Broker has a strong incentive to maximize Ii. The utility value of $I_i$ ($U_{bi}$) is typically much greater than the utility value of $I_p$, ($U_{bp}$) because $I_i$ is perceptively more 'permanent', easily accessible regardless of time, and is typically more credible. In some jurisdiction, there may be rules/laws about the minimum amount of information that the broker has to disclose about a property. The tradeoff between providing too much information $I_i$, and disintermediation of the broker is substantially affected by the existence of states $E_s$, $E_m$ or $E_p$. On information and price dispersion on the Internet, see: Baylis & Perloff (2002); Arnold (2000); Bakos (1997); Buyukozkan G (2004); Bryjolfsson & Smith (1999).

Hence the real estate broker's general problem is a multi-criteria maximization and minimization problem:

$$\text{Min } [\{B_b + B_s + B_i + B_n\} \mid \text{Min }(E_s, E_p, E_m)]$$
$$\text{Max } [SC_{br} + RC_{br}]$$

OR

$$\text{Max } [SC_{br} + RC_{br} - B_b - B_s - B_i - B_n] \mid \text{Min }(E_s, E_p, E_m)]$$

s.t.:

$cP > \text{Max } [0; (B_b + B_s + B_i)]$; where $cP$ = fixed amount $x$.
$c, P > 0$
$\rho_p > 0$
$\rho_i > 0$

In calculating the Buyer's search costs, the buyer's time and the buyer's information processing capabilities are major components, and a distinction has to be made between two types of 'time':

1. "Valued time' – for which buyer can be compensated for work done.

2. "Leisure time" – for which there is no compensation. Most searches are done using Leisure Time.

Gwin's (2004:15, 11-17) contention that "……. more information will be provided on websites in countries where buyers have a high cost of search……" is not valid. In fact the opposite is more likely to occur, because search costs are directly related to internet access and level of sophistication of telecomm infrastructure. Hence, its more likely that more information will be provided on a US realtor's website, than on a Costa Rican realtor's website because:

- Internet penetration is much lower in Costa Rica.

- The cheapest search is online searches.

- The US realtor is more wiling to adapt to the Internet than the Costa Rican realtor (because the US realtor is more likely to be at a substantial competitive disadvantage by not using the Internet).

- Most standardized real estate data is delivered via the Internet.

- The US realtor's company/office is more likely to be part of a larger network of brokers rely on the Internet for daily operations.

**4. Conclusion**

Broker disintermediation can radically alter the structure of the real estate industry, and coupled with the Internet, can radically change the nature of buying and selling processes and the mortgage lending processes.

This article has shown that full broker dis-intermediation maybe optimal for the seller and or buyer under certain conditions; while partial dis-intermediation may also be optimal under certain conditions. With continued increases in internet availability, partial disintermediation is likely to become the most common form of relationship between the seller and the broker.

**5. Bibliography.**
**1.** Agarwal R, Bohner M, O'Regan D & Peterson A (2002). Dynamic Equations On Time Scales: A Survey. *Journal of Computational And Applied Mathematics*, 141:1-26.
**2.** Al-Ubaydli O (March 2005). *Formalizing Social Capital.* Working Paper, University of Chicago.
**3**. Anderson R & Fok R (1998) Efficiency of Franchising In The Residential Real Estate Brokerage Market. *Journal Of Consumer Marketing*, 15(4): 386-396.
**4.** Anderson R., Lewis R & Zumpano L (1999). Residential Real Estate Brokerage Efficiency From a Cost and Profit Perspective. *Journal of Real Estate Finance and Economics*, 20(3): 295-310.
**5.** Anglin P (1997). Determinants Of Buyer Search In A Housing Market. *Real Estate Economics*, 25(4): 568-578.
**6.** Anglin P & Arnott R (1991). Residential Real Estate Brokerage As A Principal-Agent Problem. *Journal Of Real Estate Finance & Economics*, 4(2): 99-125.
**7.** Angilini P (2005). Contracts For The Sale of Residential Real Estate. *Journal of Real Estate Finance & Economics,* 8(3): 195-211.
**8.** Annen K (_____). Social Capital Within The Urban Small-Firm Sector In Developing Countries: A Form Of Modern Organization Or A Reason For Economic Backwardness ? Working Paper, Washington University, St. Louis, USA.
**9**. Arnold M (1999). Search, Bargaining And Optimal Asking Prices. *Real Estate Economics*, 27(3):453-481.
**10.** Arthur B (April 1999). Complexity And The Economy. *Science*, 284(2): 107-109.
**11.** Austin, A.D.(1973). The Antitrust Threat to Real Estate Brokerage. *Real Estate Review,* 2: 9-14.
**12.** Bajtelsmit V & Worzala E (1997). Adversarial brokerage in residential real estate transactions: The Impact Of Separate Buyer Representation. *The Journal of Real Estate Research*, 14(1/2):65-75.
**13.** Bakos J (1997). Reducing Buyer Search Costs: Implications For Electronic Market Places. *Management Science*, 43:1676-
**14.** Bakos Y, Lucas H, Oh W & Simon G et al (2005). The Impact of E-Commerce on Competition in the Retail Brokerage Industry. *Information Systems Research*, 16(4):352-371,433-434.
**15.** Barlett, R.(1981). Property Rights and the Pricing of Real Estate Brokerage. *The Journal of Industrial Economics*, 30: 79-94.
**16.** Baylis K & Perloff J (2002). Price Dispersion On The Internet: Good Firms And Bad Firms. *Review Of Industrial Organization*, 21:305-324.

**17.** Beer R (2000).  Dynamic Approaches To Cognitive Science.  *Trends In Cognitive Sciences*, 4(3):91-94.
**18.** Benjamin J, Chinloy P, Jud D & Winkler D (2006).  Franchising In Residential Brokerage.  *Journal Of Real Estate Research*, 28(1): 61-67.
**19.** Benjamin J. D.,Jud  G. D. & Sirmans S (2000). What Do We Know about Real Estate Brokerage*?.  Journal of Real Estate Research*, 20(½):5-30.
**20.** Benjamin J, Chinloy P, Judd D & Winkler D (2005).  Technology and Real Estate Brokerage Financial Performance. *Journal of Real Estate Research*, 27(4): 409-426.
**21.** Black E, Carnes T & Richardson V (2000).  The Market Valuation Of Corporate Reputation.  *Corporate Reputation Review*,  .
**22.** Bond M &, Seiler M, Blake B (2000).  Uses Of Websites For Effective Real Estate Marketing.  *Real Estate Portfolio Management,* 6(2):203-210.
**23.** Bouzdine T & Bourakova-Lorgnier M (April 2004).  *The Role of Social Capital Within Business networks: Analysis Of Structural And Relational Arguments.*  Prepared for The Fifth European Conference On Organizational Knowledge, Learning And Capabilities, April 5-6, 2004, Innsbruck, Austria.
**24**. Bowles S & Gintis H (2002).  Social Capital And Community Governance. *The Economic Journal,* 112:419-436.
**25.** Bromiley P & Papenhausen C (______).  Assumptions Of Rationality And Equilibrium In Strategy Research: The Limits of Traditional Economic Analysis.  *Strategic Organization*, 1(4): 413-437.
**26.** Burt R (2000).  The Network Structure Of Social Capital, in Sutton R & Staw B (eds.), *Research In Organizational Behavior* (JAI Press, Greenwich, USA).
**27.** Bryjolfsson E & Smith M (1999).  *Frictionless Commerce ? A Comparison Of Internet And Conventional Retailers.*  Working Paper, Sloan School Of Management, MIT, USA.
**28.** Bui T, Yen J & Sankaran S (2001).  A Multi-Attribute Negotiation Support System With Market Signaling For Electronic Markets.  *Group Decision & Negotiation*, 10: 515-537.
**29.** Buyukozkan G (2004).  Multicriteria Decision making For E-Marketplace Selection.  *Internet Research*, 14(2):139-154.
**30**. Chemmanur T & Fulghieri P (1994).  Investment bank Reputation, Information production And Financial Intermediation.  *Journal Of Finance*, 55:1105-1131.
**31**. Colman A (2003).  Cooperation, Psychological Game Theory And Limitations Of Rationality In Social Interaction.  *Behavioral And Brain Sciences*, 26:139-198.
**32.** Curran C & Schrag J (2000).  Does It Matter Whom An Agent Serves: Evidence From Recent Changes In Real Estate Agency Law.  *Journal Of Law & Economics*, 43(1): 236-264.
**33.** Dale-Johnson D (1998).  Housing Market Conditions, Listing Choice And MLS Market Share.  *Real Estate Economics*, 26(2): 10-20.
**34**. Dasgupta P (2005).  Economics Of Social Capital.  *The Economic Record*, 81(255): 2-21.
**35.** Delcoure N & Miller (2002).  International Residential Real Estate Brokerage Fees And Implications For The US Brokerage Industry.  *International Real Estate Review*, 5(1): 12-39.
**36.** Dellnitz M & Junge O (1999).  On The Approximation Of Complicated Dynamical Behavior.  *SIAM Journal Of Numerical Analysis*, 36(2):491-515.
**37.** Diehl R, Kornish L & Lynch J (2003).  Smart Agents: When Lower Search Costs For Quality Information Increase Price Sensitivity.  *Journal Of Consumer Research,* 30(1):56-71.
**38.** Dolde W & Tirtiroglu D (1997).  Temporal and spatial information diffusion in real estate price changes and variances.  *Real Estate Economics*, 25(4): 539-565.
**39**. Dotzour, M. G., Morehead E & Winkler D. (1988).  The Impact of Auctions on Residential Sales Prices in New Zealand*. Journal Of Real Estate Research*, 16(1): 57-71.
**40.** Durlauf S (2002).  On The empirics Of Social Capital.  *The Economic Journal*, 112: 459-469.
**41.** D'Urso V. (______).  Internet Use And The Duration Of Buying And Selling In The Residential Housing Market: Economic Incentives And Voting.  PHD Thesis, Department of Economics, MIT.
**42.**  Elder H (2000). Buyer Brokers: Do They Make A Difference ?: Their Influence On Selling Price And Search Duration.  *Real Estate Economics*, 28(2): 1-10.
**43.** Evans L (1998). Partial Differential Equations. *Bulletin of The American Mathematical Society*, 37(3):363-367.

**44.** Ferrary M (2003). Trust And Social Capital In The Regulation Of Lending Activities. *Journal of Socio-economics*, 31:673-699.
**45.** Fang L (2005). Investment Bank Reputation And The price And Quality Of Underwriting Services. *Journal of Finance*, LX(6): 2729-2739.
**46.** Fine B (1999). The Development State Is Dead – Long Live Social Capital ? *Development & Change*, 30:1-19.
**47**. Fleck R. (2000). When Should Market-Supporting Institutions Be Established ?. *Journal Of Law, Economics And Organization*, 16: 129-154.
**48.** Friedman Y & Sandler U (1996). Evolution Of Systems Under Fuzzy Dynamic Laws. *Fuzzy Sets And Systems*, 84:61-72.
**49.** Garmaise M & Moskowitz T (2004). Confronting Information Asymmetries: Evidence From Real Estate Markets. *Review Of Financial Studies*, 18(2): 405-437.
**50.** Geltner D, Kluger B & Miller N (1991). Optimal Price And Selling effort From The perspectives Of The Broker And Seller. *AREUEA Journal*, 19(1): 1-10.
**51.** Glaeser E & Laibson D & Sacedote B (2002). The Economic Approach To Social Capital. *Economic Journal*, 112: 437-458.
**52.** Guennif S & Revest V (2005). Social Structure And Reputation: The NASDAQ Case Study. *Socio-Economic Review*, 3:417-436.
**53.** Guttery R. S., Baen J & Benjamin J. (2000). Alamo Realty: The Effects of Technology Changes on Real Estate Brokerage. *Journal of Real Estate Practice and Education*, 3(1):71-84.
**54.** Gwin C (2004). International Comparisons of Real Estate E-information on the Internet. *The Journal Of Real Estate Research*, 26(1): 1-23.
**57.** Hellman T & Murdock K (1998), Financial Sector Development Policy: The Importance Of Reputational Capital And Governance". In "Development Strategy And Management Of The Market Economy", Volume 2, eds., R Sabot & I Skekely (Oxford University press, 1998).
**58** Hertwig R & Ortmann A (2001). Experimental Practices In Economics: A Methodological Challenge For Psychologists ? *Behavior & Brain Sciences*, 24:383-451.
**59** Hodgkinson G (1997). Cognitive Inertia In A Turbulent Market: The Case Of UK Residential Estate Agents. *Journal Of Management Studies,* 34(6): 921-925.
**60.** Iacus S (2001). Efficient Estimation Of Dynamical Systems. *Nonlinear Dynamics And Econometrics*, 4(4):213-226.
**61**. Iri M (1997). Roles Of Automatic Differentiation In Nonlinear Analysis And high-Quality Computation. *Nonlinear Analysis: Theory, Methods, & Applications*, 30(7): 4317-4328.
**62.** Izmailov A & Solodov M (2001). Optimality Conditions For Irregular Inequality-Constrained Problems. *SIAM Journal of Control & optimization*, 40(4):1286-1290.
**63.** Janssen C. T. L. & Jobson J D (1980). On The Choice Of Realtor. *Decision Sciences*, 11:299-311.
**64.** Johnson K, Springer T & Brockman C (2005). Price Effects Of Non-Traditionally Broker-Marketed Properties. *Journal Of Real Estate Finance & Economics*, 31(3): 332-341.
**65.** Jud G. D., Rogers R & Crellin G. (1994). Franchising And Real Estate Brokerage. *Journal of Real Estate Finance and Economics*, 8(1):87-93.
**66**. Jud, G. D. & Winkler D. (1998). The Earnings of Real Estate Salespersons and Others in the Financial Services Industry. *Journal of Real Estate Finance & Economics*, 17(3):279-291.
**67.** Jud G. D. & Winkler D. (2000). A Note on Licensing and the Market For Real Estate Agents. *Journal of Real Estate Finance and Economics*, 21(2): .
**68.** Kauffman R, Subramani M & Wood C (2000). *Analyzing Information Intermediaries In Electronic Brokerage*. Proceedings Of The 33rd Hawaii International Conference On System Sciences,
**69.** Korczynski M & Ott U (2005). Sales Work Under Marketization: The Social Relations Of The Cash nexus ? *Organization Studies*, 26(5): 707-728.
**70.** Kreps D & Wilson R (1982). Reputation And Imperfect Information. *Journal Of Economic Theory*, 27:253-279.
**71.** Lee B, Y W, Maguluru N & Nichols M (2006). Enhancing Business Networks Using Social Networks Based On Virtual Communities. *Industrial Management & Data Systems*, 106(1): 121-138.

**72.** Lewis, D. and R. Anderson. (1999). Is Franchising More Cost Efficient? The Case of the Residential Real Estate Brokerage Industry. *Journal of Real Estate Economics*, 27(3): 545-560.
**73.** Lewis D & Anderson R (1999). Residential Real Estate Brokerage Efficiency And The Implications Of Franchising: A Bayesian Approach. *Real Estate Economics*, 27(3): 543-560.
**74.** Lurie L (2004). Decision Making In Information Rich Environments: The Role Of Information Structure. *Journal Of Consumer Research*, .
**75.** Mantrala S & Zabel E (1995). The Housing Market And Real Estate Brokers. *Real Estate Economics*, 23(2):161-181.
**76.** Marsh G. A. & Zumpano L (1988). Agency Theory And The Changing Role of the Real Estate Broker: Conflicts and Possible Solutions. *Journal Of Real Estate Research*, 3(2): 151-165.
**77.** Miceli T, Pancak K & Sirmans C (2000). Restructuring agency relationships in the real estate brokerage industry: An economic analysis. The Journal Of Real Estate Research, 20(1/2): 31-47.
**78.** Milgrom P & Roberts J (1982). Predation, Reputation And Entry Deterrence. *Journal of Economic Theory*, 27:280-313.
**80.** Miller, N.G. & Shedd. P (1979). Do Antitrust Laws Apply To the Real Estate Brokerage Industry? *American Business Law Journal*, 17/3, Fall.
**81.** Moore C (1991). Generalized Shifts: Unpredictability And Un-Decidability In Dynamical Systems. Nonlinearity, 4:1990230.
**82.** Mordukhovich B & Shao Y (1997). Fuzzy Calculus For Coderivatives Of Multifunctions. *Nonlinear Analysis: Theory, Methods, & Applications*, 29(6):605-626.
**83.** Muhanna W A.(2000). E-Commerce In The Real Estate Brokerage Industry. *Journal of Real Estate Practice and Education,* 3(1):1-16.
**84.** Muhanna W. & Wolf J. (2002). The impact of e-commerce on the real estate industry: Baen and Guttery revisited. *Journal Of Real Estate Portfolio Management*, 8(2):141-152.
**85.** Munneke H & Yavas A (2001). Incentives And Performance In Real Estate Brokerage. *Journal of Real Estate Finance and Economics*, 22(1):5-21.
**86.** Narayan D & Pritchett L (1999). Cents And Sociability: Household Income And Social Capital In Rural Tanzania. *Economic Development & Social Change*, 47(4):871-897.
**87.** Nelles O (2002). Non-Linear System identification. *Measurement Science & Technology*, 13:646-648.
**88.** Pollock G (2004). The Benefits And Costs Of Underwriter's Social Capital In The US Initial Public Offerings market. *Strategic Organization*, 2(4): 357-388.
**89.** Quan D (2002). Market Mechanism Choice and Real Estate Disposition: Search Versus Auction. *Real Estate Economics,* 30(3):365-384.
**90.** Quint T & Wako J (2004). On House-Swapping, The Strict Core, Segmentation And Linear Programming. *Mathematics Of Operations Research,* 29(4):861-877.
**91.** Rangan S (2000). The problem Of Search And Deliberation In Economic Action: When Social Networks Really Matter. *The Academy Of Management Review*, 25(4): 813-828.
**92.** Rutherford R, Springer T & Yavas A (2004). The Impact Of Contract Type On Broker Performance: Submarket Effects. *Journal Of Real Estate Research*, 26(3): 277-287.
**93.** Schmid A (2000). Affinity As Social Capital: Its Role In Development. *Journal Of Socio-Economics*, 29:159-171.
**94.** Schmitz S (2000). The Effects OF Electronic Commerce On The Structure OF Intermediation. *Journal Of Computer Mediated Communication*, 5(3): .
**95.** Scott J (2000). Emerging Patterns From The Dynamic Capabilities Of Internet Intermediaries. *Journal Of Computer Mediated Communication*, 5(3): .
**96.** Seierup S (1996). "Small Town Entrepreneurs And Networks In Kenya", pp. 81-99, in McCormick D & Pedersen P (eds.), *Small enterprise: Flexibility And Networking In An African Context* (Longhorn, Nairobi/Kenya).
**97.** Sirmans, C. F., Turnbull G & Benjamin J (1991). The Markets For Housing And Real Estate Broker Services. *Journal Of Housing Economics*, 1:207-217.
**98.** Stavrovski B (2004). Designing new e-business model for a commercial real estate enterprise: a case study. *Online Information Review*, 28(2):110-119.
**99.** Swain N (2003). Social Capital And Its Uses. *Archives Of European Sociology*, 2:185-212.

**100.** Tse R & Webb J (2002). The effectiveness of a Web strategy for real Estate brokerage. *Journal Of Real Estate Literature*, 10(1):121-130.
**101.** Tuccillo J (1997). Technology And The Housing Markets. *Business Economics,* 32(3):17-20.
**102.** Turnbull, G. K. (1996). Real Estate Brokers, Nonprice Competition And The Housing Market. *Real Estate Economics,* 24(3): 293-304.
**103.** Van Dalen H (June 1995). *Efficiency And Collusion In Dutch Real Estate Brokerage: The Case Of The Twentieth Century Middlemen's Guild.* Working Paper, Research Center For Economic Policy (OCFEB), Erasmus University, Netherlands.
**104.** Van Gelder T (1998). The Dynamical Hypothesis In Cognitive Science. *Behavioral And Brain Sciences*, 21:615-665.
**105.** Wachter S.M. (1987). *Residential Real Estate Brokerage: Rate Uniformality and Moral Hazard*. Research In Law and Economics series, Greenwich, CT and London, JAI Press, 10: 189-210.
**106.** Wheaton W (1990). Vacancy, Search, And Prices In A Housing Market Matching Model. *The Journal of Political Economy*, 98(6):1270-1290.
**107**. Williams J (1998). Agency And Brokerage of Real Assets In Competitive Equilibrium. *Review Of Financial Studies,* 11(2): 29-249.
**108.** Williams J (1998). Agency And Brokerage Of Real Assets In Competitive Equilibrium. *Review Of Financial Studies*, 11(2): 239-280.
**109.** Worzala E.M. & McCarthy M (2006). Landlords, Tenants and E-commerce: Will the Retail Industry Change Significantly?, *Journal of Real Estate Literature*, forthcoming.
**110.** Yang, T. L., Trefzger J & Sherman L F (1997). A Microeconomic Study Of Commercial Real Estate Brokerage Firms*. Journal of Real Estate Research*, 13:177-194.
**111.** Yavas A (1992). A simple search and bargaining model of real estate markets. *Journal of the American Real Estate and Urban Economics*, 20(4):533-550.
**112.** Yavas A. (2001). Impossibility Of A Competitive Equilibrium In The Real Estate Brokerage Industry. *Journal of Real Estate Research,* 21(3)::187-200.
**113.** Yavas A & Colwell P (1999). Buyer Brokerage: Incentive And Efficiency Implications. *Journal of Real Estate Finance & Economics*, 18(3): 259-277.
**114.** Yavas A, Miceli T & Sirmans C (2001). An Experimental Analysis Of The Impact Of Intermediaries On The Outcome Of Bargaining Games. *Real Estate Economics*, 29(2):251-276.
**115.** Yinger J. (1993). A Search Model Of Real Estate Broker Behavior. *American Economic Review*, 71(4): 591-604.
**116.** Wu C & Colwell P (1986). Equilibrium Of Housing And Real Estate Brokerage Markets Under Uncertainty. *AREUEA Journal*, 14(1):1-23.
**117.** Zumpano, L. V. & Elder H (1994). Economies Of Scope And Density In The Market for Real Estate Brokerage Services. *AREUEA Journal*, 22:497-513.
**118.** Zumpano L, Elder L, Elder H & Crellin G (1993). The Market for Residential Real Estate Brokerage Services: Costs Of Production And Economies of Scale. *Journal of Real Estate Finance and Economics*, 6(3):237-250.
**119.** Zumpano, L. V., Elder H & Baryla E (1996). Buying a House And The Decision To Use A Real Estate Broker. *Journal Of Real Estate Finance & Economics*, 13:169-181.
**120**. Zumpano L. V., Anderson R, Baryla A E & Johnson K (2001). Internet Use and Real Estate Brokerage Market Intermediation. Working paper presented at the American Real Estate Society Annual Meeting, Coeur d'Alene, Idaho, USA; April 2001.